\documentstyle[12pt]{article}

\begin{document}

%\vskip 0.5 cm
%\vskip 1.5 cm

\begin{flushleft}
{\large\bf Proposal of a second generation of quantum-gravity-motivated
Lorentz-symmetry tests:
\smallskip sensitivity
to effects suppressed quadratically by the Planck scale\footnote{This essay
received an ``honorable mention" in the 2003 Essay Competition of the Gravity
Research Foundation.}
}
%{\large\bf The next goal of quantum-gravity experiments:\\
%sensitivity to effects suppressed quadratically by the Planck scale
%}
\end{flushleft}

\vskip 0.5 cm
\begin{center}
{\bf Giovanni AMELINO-CAMELIA}\\
\end{center}

\begin{center}
{\it Dip.~Fisica Univ.~Roma ``La Sapienza''
and Sez.~Roma1 INFN, }\\
{\it Piazzale Moro 2, 00185 Roma, Italy}
\end{center}

\vskip 0.5 cm
%\vspace{1cm}
\begin{center}
{\bf ABSTRACT}
\end{center}

\begin{quotation}
\leftskip=0.6in \rightskip=0.6in

{\small
\noindent
Over the last few years the study of possible Planck-scale
departures from classical Lorentz symmetry has been one
of the most active areas of quantum-gravity
research. We now have a satisfactory description
of the fate of Lorentz symmetry in the most popular
noncommutative spacetimes and several studies have been devoted
to the fate of Lorentz symmetry in loop quantum gravity.
Remarkably
there are planned experiments with enough sensitivity
to reveal these quantum-spacetime
effects, if their magnitude is
only linearly suppressed by the Planck length.
Unfortunately, in some quantum-gravity scenarios
even the strongest quantum-spacetime effects are suppressed by
at least two powers of the Planck length, and many authors have argued
that it would be impossible to test
these quadratically-suppressed effects.
I here observe that advanced
cosmic-ray observatories
and neutrino observatories
can provide the first elements of
an experimental programme testing the possibility of
departures from Lorentz symmetry that are quadratically Planck-length
suppressed.}
\end{quotation}

\baselineskip 17.5pt plus .5pt minus .5pt

The recent interest in the possibility that (classical)
Lorentz symmetry might be only an approximate symmetry of
quantum spacetime was originally ignited by phenomenological
analyses based on mostly-heuristic
arguments~\cite{grbgac,kifuaus,ita,gactpGACQM100}.
It has now matured into a detailed technical understanding of
the fate of Lorentz symmetry in quantum-gravity approaches
based on noncommutative
geometry~\cite{susskind,gacdsrGACDSRNATURE,jurekNEWkappamin}
and we also have several insightful results
concerning the loop-quantum-gravity
approach~\cite{gampul,mexweaveThiemNEWdisprel,alfaro,leeREVIEW}.

In approaches involving spacetime discretization,
such as loop quantum gravity, we are
essentially finding~\cite{gampul,mexweaveThiemNEWdisprel,alfaro,leeREVIEW}
that the discretized spacetime observables
are incompatible with continuous Lorentz-symmetry transformations.
Lorentz symmetry remains a good approximate symmetry, since it is violated only
by correction terms whose magnitude is governed by the ratio $L_p/\lambda$,
where $L_p \simeq 10^{-33}cm$ is the
Planck length and $\lambda$
is a much larger characteristic length scale present in the physical
context ({\it e.g.}, the wavelength of a particle).

When the quantum-gravity approach is based on noncommutative geometry
Lorentz symmetry is not necessarily broken, but it must at least
be ``deformed" in an appropriate sense~\cite{gacdsrGACDSRNATURE,jurekNEWkappamin}.
Classical
Lorentz symmetry is described through a standard Lie algebra, with an
associated ordinary description of the action of the symmetry generators
on products of fields. A deformed
action of the Lorentz generators is required by
consistency with fact that
fields in a noncommutative geometry
are themselves noncommutative. The action of symmetry generators
on products of fields requires the introduction
of additional structures
(the so-called co-algebra sector~\cite{lukie}),
effectively replacing a Lie algebra
of symmetries by a Hopf algebra
of symmetries.

There is a common prediction of these otherwise different
descriptions of the fate of Lorentz symmetry:
Planck-scale-modified dispersion relations.
These are primarily characterized, from a phenomenological
perspective, through the presence of a
lowest-$L_p$-order correction
\begin{equation}
E^2 \simeq p^2 + m^2 -  \eta (L_p E)^n p^2 ~,
\label{disp1}
\end{equation}
where $\eta$ is a coefficient of order 1, whose precise value
may depend on the specific model,
and $n$, the lowest power of $L_p$ that leads to a nonvanishing
contribution, is also model-dependent.
In any given noncommutative geometry
one finds a definite value of $n$, and it appears to be equally
easy~\cite{gacdsrGACDSRNATURE,jurekNEWkappamin}
to construct noncommutative geometries with $n=1$ or with $n=2$.
In loop quantum gravity one might typically
expect to find $n=2$, but certain scenarios~\cite{gampul}
have been shown to lead to $n=1$.

The difference between $n \! = \! 1$ and $n \! = \! 2$
is very significant from
a phenomenological perspective. Already with $n \! = \! 1$,
which corresponds to effects that are linearly suppressed by the Planck
length, the correction
term in Eq.~(\ref{disp1}) is very small: assuming $\eta \! \simeq \! 1$,
for particles with energy  $E \sim 10^{12} eV$,
the highest-energy particles we produce in laboratory,
it represents a correction of one part in $10^{16}$.
%, and even for the
%highest-energy particles we observe in astrophysics ($E \sim 10^{20} eV$)
%it is just a correction of one part in $10^8$.
Of course, the case $n \! = \! 2$
pays the even higher price of quadratic suppression by the Planck length
and for $E \! \sim \! 10^{12} eV$ its effects are at the  $10^{-32}$
level.
% and for the highest-energy particles we observe in astrophysics
% its effects are at the $10^{-16}$ level.

The realization that quantum-gravity effects
(whether or not they involve departures
from classical Lorentz symmetry)
are inevitably suppressed by the smallness of the Planck length
generated for many decades the conviction~\cite{ishamREVIEW}
that no guidance from experiments could be obtained in the
study of the quantum-gravity problem.
Recently this pessimistic view was
revised~\cite{leeREVIEW,carloMGcarlipREVIEW}
thanks to the results obtained in the mentioned Lorentz-symmetry
studies~\cite{grbgac,kifuaus,ita,gactpGACQM100,gacdsrGACDSRNATURE,jurekNEWkappamin,gampul,mexweaveThiemNEWdisprel,alfaro}
and in other quantum-gravity studies not
directly connected
with the fate of Lorentz symmetry~\cite{mynapapPOLONPAP,nggwi,ahluREVIEW}.
These studies showed that in the case of effects with linear dependence
on the Planck length there is a handful
of experiments that allow testing.

For the case of Planck-scale
departures from Lorentz symmetry,
there are two well-established physical contexts in which the
predicted effects, for $n=1$,
are observably large.
The first context is the one of observations of gamma-ray bursts.
These are far-away bursts of photons
which we plan to observe soon up to $\sim TeV$ energies,
with observatories such as GLAST~\cite{glast}.
% with good statistics
According to Eq.~(\ref{disp1}) these (nearly-)simultaneously~\cite{grbgac}
emitted photons
should reach our observatories with small energy-dependent time-of-arrival
differences
\begin{equation}
\Delta t \sim \eta (L_p  E)^n T \ ,
\label{deltatime}
\end{equation}
where $T$ is the overall time of travel and $E$ is the highest
energy among the two photons whose times of arrival are being compared.
On the basis of $v = dE/dp$, Eq.~(\ref{disp1}) predicts a small energy
dependence~\cite{grbgac,kifuaus,ita,gactpGACQM100,gacdsrGACDSRNATURE,jurekNEWkappamin,gampul,mexweaveThiemNEWdisprel,alfaro}
of the speed of massless particles,
and, although the energy-dependent term is very small,
for gamma-ray bursters at distances as large as $10^{10} light~years$
the time-of-arrival
differences, resulting from the whole long journey, could be
at the observable level of $0.01 s$, in the case of observations
of $GeV$ particles and $n=1$.
The analogous prediction for the case of quadratic suppression
by the Planck length ($n=2$) leads to time-of-arrival
differences of order $10^{-18} s$, which is instead
much beyond  the achievable sensitivities.

The differences between the $n \! = \! 1$ and $n \! = \! 2$
scenarios are also important in the other physical context
which is being considered as a possible way to test the
idea of Planck-scale departures from Lorentz symmetry.
This is the context of observations
of the highest-energy cosmic rays.
A characteristic feature of the expected cosmic-ray spectrum,
the so-called ``GZK limit", depends on the evaluation of the
minimum energy required of a cosmic ray in order to produce pions
in collisions with CMBR (cosmic microwave background radiation)
photons.
According to ordinary Lorentz symmetry this threshold energy
is $E_{th}  \! \simeq  \! 5 {\cdot} 10^{19} eV$ and cosmic rays with energy in
excess of this value should loose the excess
energy through pion production.
The value of a threshold energy is obtained combining
energy-momentum conservation and dispersion relations.
The modification (\ref{disp1})
of the dispersion relation could of course affect the evaluation
of the cosmic-ray threshold energy, depending on the type of laws
of energy-momentum conservation which are implemented.
In a large number of quantum-gravity studies that adopt (\ref{disp1})
it is assumed that
energy-momentum conservation is not modified, which of course comes
at the cost of obtaining results for the threshold energies that
reflect the existence of a preferred class of inertial observers.
Several recent quantum-gravity studies have also explored
the idea~\cite{gacdsrGACDSRNATURE,jurekNEWkappamin,leeDSRprd}
of implementing a dispersion relation of type (\ref{disp1})
without giving rise to a preferred class of inertial observers,
which then necessarilly requires~\cite{gacdsrGACDSRNATURE}
modified laws of boost transformations between inertial observers
and a deformed law of energy-momentum conservation.
For simplicity I focus here on the much-studied case with unmodified
energy-momentum conservation, in which
it has been shown~\cite{kifuaus,ita,gactpGACQM100,alfaro}
that Eq.~(\ref{disp1}) induces a Planck-length-dependent
contribution to the threshold energy of order
\begin{equation}
\Delta E_{th} \sim L_p^n E_{th}^{n+2}/\epsilon \ ,
\label{deltaeth}
\end{equation}
where $\epsilon$ is a representative (very low)
energy scale of the CMBR.
Strong interest was generated by the
observation~\cite{kifuaus,ita,gactpGACQM100,alfaro}
that for $n= 1$ one finds $\Delta E_{th} \gg E_{th}$,
meaning that the value of the threshold energy is
very significantly
affected by this class of quantum-spacetime effects.
This possibility actually receives encouragement by the (preliminary)
AGASA~\cite{agasa} observations of the high-energy cosmic-ray spectrum, which
appear to reflect~\cite{kifuaus,ita,gactpGACQM100,alfaro}
a sizeable shift of the standard GZK limit.

Modifications of threshold energies that are formally similar
to the one here considered occur also in the analysis of
processes we study in the laboratory, but in those contexts
the correction is completely negligible (essentially the
relevant analog of the ratio $E_{th}/\epsilon$ is not
large enough).
Instead in collisions involving a ultra-high-energy cosmic-ray
and a CMBR photon the ratio $E_{th}/\epsilon$ is
large enough to compensate for the smallness of the
Planck length, at least if $n = 1$.
The case $n=2$ is usually not considered in the cosmic-ray
literature because of the large additional suppression
introduced by the extra power of $L_p$.

The fact that at least for $n=1$ these two experimental strategies
(time-of-arrival analyses of gamma-ray bursts and
cosmic-ray spectrum analyses) allow testing effects that are genuinely
at the Planck scale has generated very strong interest,
as one can infer from
recent quantum-gravity reviews~\cite{leeREVIEW,carloMGcarlipREVIEW}.
But, as discussed above, a more satisfactory exploration of this idea of
quantum-gravity-induced departures from Lorentz symmetry should
also consider the case of quadratic Planck-length
suppression ($n=2$),
and it is generally believed~\cite{leeREVIEW,carloMGcarlipREVIEW}
that it will never be possible to find contexts with
sensitivity to quadratically-$L_p$-suppressed effects.

The key point that I intend to convey here is that, contrary to
these common beliefs, we might soon be ready for
experimental studies with good sensitivity even to effects
that are quadratically suppressed by the Planck length.
This can be achieved even just reconsidering  the
mentioned time-of-arrival analyses and cosmic-ray spectrum analyses.

Let me start by reconsidering
the analysis of the cosmic-ray spectrum.
A large majority of related quantum-gravity studies
focus on the fact that for $n=1$
one finds $\Delta E_{th} \gg E_{th}$,
which would render possible~\cite{kifuaus,ita,gactpGACQM100,alfaro}
the observation of cosmic rays
much above the GZK limit of $\sim 5 {\cdot} 10^{19} eV$
that one obtains on the basis of a standard estimate of $E_{th}$.
The possibility to constrain the case $n=2$ was already considered
in Ref.~\cite{ita} and a possible role of the case $n=2$ in
the cosmic-ray paradox
was investigated in detail in Ref.~\cite{gactpGACQM100}, but
these studies went largely unnoticed in this respect.
Actually, even for $n=2$  one finds that $\Delta E_{th} \sim E_{th}$, {\it i.e.}
the quantum-gravity correction is nonnegligible.
This is easily verified by substituting the relevant energy
scales in relation (\ref{deltaeth}) for $n=2$.
While in the $n=1$ case the effect is very large,
the case $n=2$ invites us to consider the case of a $\Delta E_{th}$
which is comparable to
(but not necessarily much larger than) $E_{th}$, and this could
in turn require cosmic-ray observations with high-statistics
information on the
spectrum at energies close to the GZK limit.
Since the Planck-scale $n=2$ correction $\Delta E_{th}$ is not
negligible,
such detailed studies of the spectrum in the neighbourhood
of $E_{th}$ would inevitably provide an opportunity for a significant
test. Cosmic-ray observations such as
the ones planned for the Auger observatory~\cite{auger}
do have the required
capability for high-statistics studies of cosmic rays with
energies close to $E_{th}$.

The corresponding ``$n=2$ upgrade" of the time-of-arrival
gamma-ray-burst studies actually requires an even more
profound modification of the experimental strategy.
As one can easily deduce from (\ref{deltatime}), in order to
compensate for the extra power of $L_p$ which is present
in the case $n=2$  it would be necessary to
compare the times of arrival of ultra-energetic particles emitted by
a gamma-ray burster.
Unfortunately, we only expect to
observe gamma-ray-burst photons with energies up to the $TeV$ scale,
since at higher energies
photons are efficiently absorbed before they reach the Earth.
However current well-established models~\cite{grbNEUTRINOnew} predict
that gamma-ray bursters should also emit a substantial amount of
high-energy neutrinos.
With advanced planned neutrino observatories, such as
ANTARES~\cite{antares}, NEMO~\cite{nemo} and EUSO~\cite{euso},
it should be possible to correlate detections of
high-energy neutrinos with corresponding detections
of gamma-ray-burst photons.
Neutrinos are of course immune
from electromagnetic absorption,
and therefore
it is possible to observe neutrinos with energies
between $10^{14}$ and $10^{19}$ $eV$.

Models of gamma-ray bursters predict in particular a substantial
flux of neutrinos with energies of about $10^{14}$ or $10^{15}$ $eV$.
Comparing the times of arrival of these neutrinos emitted by
gamma-ray bursters to the corresponding times of arrival of
low-energy photons the case $n=1$ would predict a
huge\footnote{The strong implications
of the case $n=1$ for neutrino astrophysics
were already emphasized in Refs.~\cite{grbNUnick,grbNUqg}.
Ref.~\cite{grbNUqg} did not at all consider the case $n=2$, on which
I focus here, while Ref.~\cite{grbNUnick} did make a brief remark
(see footnote 6) one the case $n=2$, stressing that,
according to the scheme there advocated, one could find
no motivation for the case $n=2$
but in principle good sensitivity might be achieved
through observations of high-energy neutrinos.}
time-of-arrival difference ($\Delta t \sim 1 year$)
and even for the case $n=2$ the expected time-of-arrival
difference, $\Delta t \sim 10^{-6} s $, is within the realm
of possibilities of future observatories.

Current models of gamma-ray bursters also predict some
production of neutrinos
with energies extending to the $10^{19} eV$ level.
For such ultra-energetic neutrinos a comparison of time-of-arrival
differences with respect to soft photons also emitted by the burster
should provide, assuming $n \! = \! 2$,
a signal at the level $\Delta t \! \sim \! 1 s $, comfortably within
the realm of timing accuracy of the relevant
observatories.
For this strategy relying on ultra-high-energy neutrinos
the delicate point is clearly not timing, but rather the statistics
(sufficient number of observed neutrinos)
needed to establish a robust experimental result.
Moreover, it appears necessary to understand gamma-ray bursters
well enough to establish if there is a typical time delay
after a gamma-ray-burst low-energy-photon trigger at which we should
expect the arrival of neutrinos. The fact that this ``time history"
of the gamma-ray burst must be obtained only with precision
of, say, $1 s $ (which is a comfortably large time scale with respect
to the short time scales present in most gamma-ray bursts)
suggests that this understanding should be achievable in the short-term
future. Hopefully the observations I am reporting here
will provide additional
motivation for these studies.

In summary, the next few years,
following the strategies proposed
here, should mark the beginning of
a second generation of quantum-gravity-motivated
Lorentz-symmetry studies, in which
we explore even the possibility of
effects suppressed quadratically by the Planck length.
High-energy neutrino observatories can lead to very significant
insight, which is particularly valuable since the time-of-arrival
studies are not affected by the law of energy-momentum conservation,
and are therefore applicable to studies of Planck-scale
effects of type (\ref{disp1}) with and without the emergence of
a preferred class of inertial observers.
As mentioned, a role for high-energy neutrino observations
in quantum-gravity research requires some progress particularly
in our understanding of gamma-ray bursts, and it is therefore
an objective which we can realistically set presently as a goal
and expect to achieve in a few years.
Forthcoming improvements~\cite{auger} in our knowledge of ultrahigh-energy
cosmic rays are likely to be our first step into
the realm of  quadratically Planck-length-suppressed effects,
but they will provide us indications that are applicable to a more
limited class of quantum-gravity models, since threshold analyses
are affected
by the delicate issue of a possible role for the Planck length in
the laws of energy-momentum conservation.

\section*{Acknowledgments}
I am grateful for informative conversations
on the expected advances of neutrino astrophysics with A.~Capone and T.~Piran.

\vfil
\eject

\end{document}